\documentclass[11pt]{article}

\usepackage{authblk} 

\usepackage[T1]{fontenc}
\usepackage[utf8]{inputenc}
\usepackage{newtxtext}
\usepackage{newtxmath}
\usepackage{microtype}
\usepackage{booktabs}

\usepackage{geometry}

\usepackage{graphicx}
\usepackage{hyperref}
\usepackage{caption}

\usepackage{tikz}
\usetikzlibrary{arrows, decorations.pathreplacing}

\usepackage{natbib}
\newcommand{\qname}[1]{\texttt{#1}}
\usepackage[ttdefault=true]{AnonymousPro}
\usepackage{float}

\usepackage{setspace}

\makeatletter
\newif\if@anonymize
\@anonymizefalse

\newcommand{\blinded}[2]{%
\if@anonymize
#2%
\else
#1%
\fi}

\title{Think-aloud interviews: A tool for exploring student statistical reasoning}

\if@anonymize
\author{(Blinded authors)}
\else
\author[1]{Alex Reinhart}
\author[2]{Ciaran Evans}
\author[3]{Amanda Luby}
\author[4]{Josue Orellana}
\author[1]{Mikaela Meyer}
\author[5]{Jerzy Wieczorek}
\author[1]{Peter Elliott}
\author[1]{Philipp Burckhardt}
\author[1]{Rebecca Nugent}
\affil[1]{Department of Statistics \& Data Science, Carnegie Mellon University}
\affil[2]{Department of Mathematics and Statistics, Wake Forest University}
\affil[3]{Department of Mathematics \& Statistics, Swarthmore College}
\affil[4]{Center for the Neural Basis of Cognition and Machine Learning Department, Carnegie Mellon University}
\affil[5]{Department of Statistics, Colby College} 

\fi
\usepackage[colorinlistoftodos]{todonotes}

\begin{document}

\maketitle

\begin{abstract}
  Think-aloud interviews have been a valuable but underused tool in statistics education research. Think-alouds, in which students narrate their reasoning in real time while solving problems, differ in important ways from other types of cognitive interviews and related education research methods. Beyond the uses already found in the statistics literature---mostly validating the wording of statistical concept inventory questions and studying student misconceptions---we suggest other possible use cases for think-alouds and summarize best-practice guidelines for designing think-aloud interview studies. Using examples from our own experiences studying the local student body for our introductory statistics courses, we illustrate how research goals should inform study-design decisions and what kinds of insights think-alouds can provide. We hope that our overview of think-alouds encourages more statistics educators and researchers to begin using this method.
\end{abstract}

\section{Introduction}

Think-aloud interviews, in which interview subjects solve problems while narrating their thinking aloud, provide a valuable statistics education research tool that can be used to study student misconceptions, improve assessments and course materials, and inform teaching. In contrast to written assessment questions or traditional interviews, think-alouds involve the subject describing their thoughts in real time without interviewer feedback, rather than providing explanations after the fact or in dialogue with the interviewer \citep{Ericsson_1998, Adams2011}. They differ in important ways from other types of cognitive interviews \citep{leighton2017using}, such as the task-based interviews in \citet{woodard_2021}, where
the interviewer may probe the interviewee about their steps or thought process as they work. Think-aloud interviews better capture how interviewees think about the problem on their own, and give a clearer picture of the reasoning process in real time. 

While think-alouds and other cognitive interviews are widely used in education research \citep{Bowen_1994,Kaczmarczyk:2010,Adams2011,Kuba:2014,Deane:2014,Taylor:2020}, their use in statistics education appears to be mostly concentrated on developing concept inventories \citep{lane2007development, jiyoon2012, ziegler2014reconceptualizing, Sabbag:2016} and several studies on student misconceptions \citep{konold1989informal, williams1999novice, lovett2001collaborative}. Furthermore, existing work in statistics does not provide extensive guidance on the think-aloud process to inform researchers interested in conducting their own interviews.

Our goal in this work is to advocate for think-aloud interviews in statistics education by 
describing details of the think-aloud process and including best practices for how interview protocols may vary for different research goals, so that interested readers have a starting point to conduct their own interviews.


In order to illustrate how the research context should drive design decisions and the interpretation of results, we use one of our own think-aloud interview studies as a concrete running example.
In this informal study, appraising the accuracy of our own beliefs about student understanding in the introductory statistics courses we have taught at \blinded{Carnegie Mellon}{our institution}, we conducted think-aloud interviews with approximately 30 students over several semesters. Questions covered a selection of introductory topics where we believed we knew the most common misconceptions held by our students.
Findings from the study showed us several areas where we were mistaken, with clear implications for how we might revise our teaching.
Furthermore, data from our early think-alouds led us to revise several ambiguous tasks to improve our later think-alouds.
The particular results we present here are not meant to generalize beyond our student population. Rather, we include them in the hope that they inspire other statistics educators and researchers to see the value of using think-alouds in their own work.

In Section 2, we describe the motivation for think-aloud interviews, and contrast them with other tools like concept inventories. Since think-alouds can be used for a variety of research goals in statistics education, we comment on how protocols may need to change to support different types of research. In Section 3 we summarize best-practice guidelines for think-aloud interviews \citep{leighton2017using}, and we describe our own think-aloud protocol in order to illustrate how these guidelines may be applied to tailor a study design to particular research goals. In Section 4, we share several findings from our think-aloud interviews to demonstrate how we interpreted these results in light of our own research goals, focused on our students and our own teaching.
Through these case studies, we emphasize how think-alouds provided new information about our students which we had not observed in traditional interactions such as class discussions, office hours, and written assignments. Although some of the misconceptions we observed had already been discussed in the statistics education literature, we had not previously been aware of them in our own students.

\section{Background on think-aloud interviews}

\subsection{Think-alouds vs.\ related methods}

There is a spectrum of ways that instructors learn how their students think. At one end of the spectrum, instructors can talk to students as part of the course: through questions in class, in after-class chats, during office hours, and in oral exams. These conversations are intended to serve the course and the students, but also provide instructors with  glimpses into student thinking. At the other end of the spectrum, there is a range of tools for more detailed research insight into student thinking, including concept inventories and several varieties of cognitive interviews.

Concept inventories are written assessments designed to cover specific concepts. Several have been designed for introductory statistics course topics, such as $p$-values and statistical significance \citep{lane2007development}, inferential reasoning \citep{jiyoon2012}, and statistical literacy \citep{ziegler2014reconceptualizing, Sabbag:2016}. Using a pre-existing concept inventory to assess student thinking has a low time cost for the instructor, since items have already been written and validated. They can also be administered en masse to students since they are typically multiple choice and can be auto-graded. However, they offer low customizability for an instructor or researcher interested in a topic not covered, and since no information is recorded on student thinking beyond their answer choice, it is hard to assess the reason behind incorrect answers---unless the questions were specifically written to detect common misconceptions and the test was validated for this purpose, as with certain items on the Reasoning about P-values and Statistical Significance (RPASS) scale \citep{lane2007development}.

Interviews with students provide richer opportunities to cover specific topics and understand student thinking. Note these interviews are distinct from oral examinations; while oral exams may be useful for assessing student understanding \citep{theobold2021oral}, associating grades with think-aloud interviews can inhibit the ability to accurately capture student thought processes, as discussed by \citet{leighton2013item, leighton2017using}. It is important in think-alouds to reassure students that the interview is non-evaluative so they are comfortable sharing their thoughts.
Informal discussions in class or in office hours can be a less-evaluative way to understand students' thinking, but such conversations are unstructured interventions primarily meant to serve the students in achieving a course's learning objectives, not to carry out structured research into what they did or did not understand before an intervention.

In the context of research, as opposed to assigning grades to students, there are several kinds of \textit{cognitive interviews} used. In one branch of cognitive interviews, the interviewer makes structured interruptions throughout the interview to ask about the volunteer's thought process. Such interviews have been termed task-based interviews \citep{woodard_2021}, verbal probing \citep{Willis_2005}, or cognitive laboratory interviews \citep{leighton2017using}. For instance, a student volunteer completes a course-related task while the interviewer prompts the student with questions about how they chose their answer, requests feedback on the difficulty of the task, or asks leading questions to help guide the student back if they go too far off track. Similar varieties of cognitive interviews are also widely used in survey instrument design to ensure survey questions are correctly interpreted and measure the intended constructs \citep{Willis_2005} or in software design to improve the usability of a interface \citep{nielsen1993mathematical}, in which cases the interviewer may explicitly solicit the volunteer's suggestions about how to improve the survey form or the software interface.
While such prompting allows the interviewer to request additional details about the volunteer's reasoning or preferences, it tends to cause subjects to report their self-reflections \textit{about} their reasoning, which may differ from the actual reasoning process they originally used.


By contrast, think-alouds are a style of cognitive interview that involve minimal dialogue with the interviewer.  
Think-aloud interviews focus specifically on interviewee reasoning without any influence from the interviewer. In a think-aloud interview, conducted privately with the interviewee and interviewer (and potentially a designated note-taker; see below)
, the interviewer asks the subject to perform a task but requests that the subject \textit{think aloud}\footnote{Although we use the term ``think aloud'' to be consistent with the literature, communication need not be verbal. The key is to use a real-time communication method, so that participants are relying on short-term working memory in narrating \textit{while} they solve the task, not reflecting on their solution afterwards. For instance, \cite{roberts2006methods} used a ``gestural think aloud protocol'' in a study with sign language users.}
while doing so, starting by reading the task aloud and narrating their entire thought process up to the conclusion \citep{Ericsson_1998, leighton2017using}. In contrast to task-based interviews or tutoring sessions, which include dialogues between the interviewer and student, in a think-aloud interview the interviewer neither gives feedback nor offers clarification until the end of the interview, other than reminders to ``Please think out loud'' if the subject falls silent \citep{Adams2011}. This provides a better evaluation of the subject's reasoning process; \cite{Ericsson_1998} suggest that when a subject explains their reasoning only after completing a task, this ``biased participants to adopt more orderly and rigorous strategies to the problems that were easier to communicate in a coherent fashion, but in turn altered the sequence of thoughts,'' while ``the course of the thought process can be inferred in considerable detail from thinking-aloud protocols.''

That said, at times it may be useful to begin the interview protocol with a ``concurrent'' think-aloud first pass through all the tasks, then conclude with a ``retrospective'' second pass in which the interviewer may probe for more detail about how the interviewee understood the tasks or explicitly request feedback about the wording of questions \citep{leighton2017using}. \cite{branch2000investigating} contrasted think-alouds with ``Think Afters'' and found that such retrospective reports omitted many of the dead ends that her participants had clearly run into---but did provide more detailed rationales for certain steps taken, especially when tasks were so complex or absorbing that think-aloud participants did not manage to express every detail in real time.
While our paper focuses on think-alouds for capturing real-time reasoning, our list of uses in Section~\ref{sec:uses} and our summary of study-design best practices in Section~\ref{sec:think-aloud-process} also draw on related examples from other cognitive interview types when appropriate.

\subsection{Uses for think-aloud interviews}\label{sec:uses}

Think-aloud interviews have been used to elicit respondent thinking in a range of fields, including software usability studies \citep{Norgaard:2006} and many areas of education research \citep{Bowen_1994,Kaczmarczyk:2010,Adams2011,Kuba:2014,Deane:2014,Taylor:2020}. Think-aloud interviews may be useful both for studying general understanding and misconceptions about statistics concepts, and for improvements in teaching at the instructor and department level. Below we describe several potential research uses for think-aloud interviews.

\subsubsection{Developing concept inventories} 

Think-alouds have been widely used to develop concept inventories in several fields, such as biology \citep{GarvinDoxas_2008,Deane:2014,Newman_2016}, chemistry \citep{Wren_2013}, physics \citep{McGinness_2016}, and computer science \citep{Kuba:2014,Porter_2019}. In statistics, several concept inventories have used think-aloud interviews or similar cognitive interviews in the development process, including the Reasoning about P-values and Statistical Significance instrument \citep{lane2007development}, the Assessment of Inferential Reasoning in Statistics \citep{jiyoon2012}, the Basic Literacy in Statistics instrument \citep{ziegler2014reconceptualizing}, and the Reasoning and Literacy Instrument \citep{Sabbag:2016}. Cognitive interview protocols for the former two instruments allowed for structured verbal probing by the interviewer, such as ``What do you think this question is asking?'' \citep{jiyoon2012}, while the latter two instruments reported using strictly think-aloud protocols.
    
In this use, interviews help inventory designers ensure that questions assess the intended concepts and are not misunderstood by students. Such interviews generally focus on changes to the question wording, not changes to the concept being tested. For instance, \cite{lane2007development} describes how interviews prompted a change in one question from the phrase ``the experiment has gone awry,'' which several students did not understand, to the clearer ``there was a calculation error,'' which helped students to focus on the statistical concept behind the question. Apart from question wording, think-aloud verbal reports can serve as ``response process validity evidence'' \citep{Sabbag:2016}, showing that respondents answer questions by using the intended response process (statistical reasoning) and not some generic test-taking strategy. This evidence can supplement other evidence for the validity of the concept inventory, including the test content, internal structure, and other types of validity evidence not directly addressed by think-alouds \citep[Chapter 11]{Jorion_2015,Bandalos:2018}.
    
Unfortunately, details on the think-aloud protocols for past statistics concept inventories are largely recorded in unpublished dissertations. We believe such details are important enough to deserve a prominent place in the published literature, reaching a wider audience.
As \cite{leighton2021rethinking} states, ``The conditions for interviews [\ldots] actually contain critical information about the quality of collected data. [\ldots] A fairly straightforward way to enhance the collection of verbal reports is to simply include much more information about all aspects of the procedures [\ldots] This would include comprehensive descriptions of the instructions given to participants, procedures for the timing of tasks, probes and strategies used to mitigate reactivity in the response processes measured.''
Furthermore, although each of these dissertations summarizes its own chosen think-aloud protocol, we are not aware of detailed discussion in the statistics education literature about general best practices for think-aloud methods or about comparisons between different approaches.

\subsubsection{Studying expert practice} 

When teaching a skill that requires expertise and experience, it may be helpful to conduct interviews with experts to understand the specific skills students need to learn. Experts often are not aware of the exact strategies they use to solve problems, and making their approaches explicit can help develop instructional materials that better teach students to reason like experts \citep{Feldon_2006}. For example, \cite{lovett2001collaborative} used think-alouds with statistics instructors to determine which skills they used to solve each problem. Members of our research group are currently applying the same idea to study expert and student reasoning about probability models \blinded{\citep{meyer2020using}}{[citation blinded]}.
    
\subsubsection{Studying student knowledge and misconceptions} 

Data from think-alouds may help to characterize how students think about a particular topic and identify misconceptions or misguided problem-solving strategies they may have. While other structured or unstructured interviews have been used for this purpose much more often in statistics education, think-alouds have appeared in the literature a few times. For example, \cite{lovett2001collaborative} used think-alouds in a data analysis activity to explore how students analyze data. \cite{konold1989informal} explored student reasoning about probability primarily with think-alouds, but also reported using a few unplanned verbal probes. \cite{williams1999novice} explored students' understanding of statistical significance with think-alouds followed by retrospective semistructured interviews. In Section \ref{sec:case-studies} below, we describe how think-aloud interviews allowed us to discover student misconceptions about sampling distributions and histograms of which we were previously unaware. 
    
While the examples above from the statistics education literature tend to focus on qualitative interpretation of verbal reports, such verbal reports could also be carefully coded and used for quantitative analysis of the data. \cite{leighton2013item} studied how a think-aloud interviewer's portrayal of their own mathematical expertise, interacting with prior student achievement and item difficulty, can account for variability in the sophistication of students' response processes on think-alouds about high school math problems.

\subsubsection{Improving course materials}

In Section \ref{sec:case-studies} below, we describe how think-alouds revealed that some of our questions were mis-aligned with the intended concept, and that some questions were confusing even if students understood the material. As the questions we used in interviews were often taken from our own course materials, think-alouds allowed us to improve these materials for future students, and incorporate common confusions more directly into teaching material. This is similar in principle to using think-alouds for studying software usability, as was done by \citet{Norgaard:2006}.
    
\subsubsection{Informing course design} When asking students questions about correlation and causation, we found that those we interviewed were often confused about when causal conclusions could be drawn, and sometimes believed confounding variables could still be present even in randomized trials. These interviews, described in Section \ref{sec:case-studies}, along with recent papers on teaching causation \citep{Cummiskey:2020, lubke2020causal}, inspired us to explore new labs and activities for teaching correlation, causation, and experimental design. While this work is still in progress, some information can be found in \blinded{our eCOTS presentation \citep{evans2020ecots}}{[citation blinded]}.

It is important to note that reasonable think-aloud protocols may differ between different use-cases. For example, developing concept inventories likely requires sufficient sample sizes to reliably assess validity, confidence that the interviewed students are representative, and careful transcription and coding of student responses. However, a smaller study may be adequate if the goal is to improve one's own courses, rather than to generalize to a broad population. In Section \ref{sec:think-aloud-process}, we discuss such study design considerations and how they depend on the research goals.

\section{The think-aloud process}
\label{sec:think-aloud-process}

To investigate student understanding in our introductory statistics courses, we conducted a think-aloud study across several semesters in 2018--2019. In this section, we summarize the main steps and general best-practices in the think-aloud process \citep[Chapters 2 and 4]{leighton2017using}. These steps are presented in chronological order and can apply to any think-aloud study. However, specific details may change in different studies, such as the length of interviews, and choices made about questions, records, and compensation. To illustrate this, we use our own think-aloud protocol as a running example throughout this section, showing how these general best-practices can be used to guide the design of an individual study.

    \subsection{Prepare research plan and resources} When think-aloud interviews are conducted for research, they are considered human subjects research. In the United States, they may be considered exempt from full Institutional Review Board (IRB) review, but this depends on the exact circumstances and institutional policies. After developing a research plan based on the following steps, but before carrying out the research, check in with your local IRB. 
    You may also wish to ensure you have funding available for recruitment incentives, recording interviews, and transcribing recordings, as well as available team members or support staff to carry out interviews and plan other logistics (such as scheduling interviews, acquiring incentives, and keeping track of consent forms or recordings). To ensure quality and consistency, interviewers may need to be formally trained by the research team or a single interviewer may conduct all interviews.

    \textbf{In our case}, our interview protocol was reviewed and classified as Exempt by the \blinded{Carnegie Mellon}{[blinded]} University Institutional Review Board (STUDY2017\_00000581). As discussed below, we decided that recordings and transcription were not necessary for our purposes. Our department's administrative staff were able to provide logistical support, and research team members were able to conduct interviews and take notes. All interviewers and note-takers were faculty or PhD students at \blinded{Carnegie Mellon}{[blinded]} at the time of the interviews and collaborated on developing the interview protocol.
    
    \subsection{Choose interview questions} Interview questions or tasks depend on the goal of the interview process. For example, when developing a concept inventory, the interview questions should consist of the draft inventory items, each designed to target learning objectives and expected problem-solving approaches or misconceptions, usually based on a review of the literature or on the experience of expert instructors. For a concept inventory or a study on a specific misconception, we recommend picking a narrow set of tasks to engage in deeply, not a broad array of topics. In some cases, it may also be useful to begin with a round of open-ended questions, and use student answers to construct distractor answers for multiple choice questions. 
    
    When choosing interview questions, it is important that they require actual problem solving skills (rather than simple memorization), and that they are not too easy or hard for the target population of interviewees. Otherwise, the think-aloud process can fail to actually capture steps in reasoning.
    \cite{jiyoon2012} used a preliminary think-aloud interview with an expert, to ensure that the developer's intended response process for each question is indeed the one used by the expert, before continuing on to study students' response processes.
    
    \textbf{In our case}, our purpose in conducting think-aloud interviews was to explore student understanding in introductory statistics at our institution, and to investigate whether our beliefs about student misconceptions were correct. We therefore drafted questions about important introductory topics such as sampling distributions, correlation, and causation. We drafted multiple-choice rather than open-answer questions because we generally wanted to check for specific misconceptions that we expected from past experiences with our own students. In Section \ref{sec:case-studies}, we describe a small selection of the questions we asked during interviews, to illustrate our reasoning for drafting these questions and how they related to specific misconceptions we had in mind.
    
    \subsection{Recruit subjects}
    
        \subsubsection{Recruitment process} Once interview questions are ready, students are invited to participate. In line with the principle that questions should neither be too easy nor too hard, and as discussed by \citet{pressley2012verbal}, the target population of subjects is often those who are still learning the material, although this may vary depending on the research goals. When developing assessment items, interviews with students who have never seen the material before could ensure that the questions are not too easy. In other situations, interviewees could include former students from past semesters if the goal is to understand how well they retain fundamental skills over time.

        However, a researcher ought to avoid recruiting from their own course: To best capture thought processes, the subject must not feel that they are being judged or evaluated, particularly by an interviewer in a position of power \citep{leighton2013item}. Human subjects research ethics also requires that subjects not feel pressured into participating. We therefore recommend the interviewer be separate from the course, and that the course instructor play no role in interviewing or recruitment besides allowing a separate recruiter to contact students in the course.

        Even if the course instructor is not involved in interviews, students may feel pressure to participate when the instructor is a member of the broader research team. In an attempt to minimize this pressure, recruiters should emphasize that no identifying information about interviewees will be shared with any course instructor, and that participation will have no impact on their grade in the course. These reassurances should be repeated at the beginning of each interview, as discussed below.
    
        \textbf{In our case}, our research team consisted of PhD students and faculty, and most team members had experience teaching introductory statistics or working with introductory students. Introductory statistics students were recruited by a member of the research team not involved in course instruction. In our first semester of interviews (Spring 2018), recruitment took place at three points in the semester chosen to align questions given in the think-aloud interviews to recent course material. The later two semesters were on compressed summer timelines, so recruitment took place only once per class. A sample recruitment script is included in our supplementary materials. Students were offered \$20 to participate and signed up with a member of the research team not involved in conducting interviews or course instruction. Every student who volunteered to participate was interviewed, including some repeat participants over the first semester of interviews. Each participant was assigned a pseudonym, which was used throughout the interview recording and data collection process. 
        
        In total, 31 students participated across three terms, resulting in 42 hour-long think-aloud interviews (33 interviews with 22 students in Spring 2018; three in Summer 2018; and six in Summer 2019).
        In Section \ref{sec:case-studies}, we focus on case studies for a subset of five questions, which were answered by 24 different students. All interviews were conducted by those members of the research group who were not teaching an introductory course at the time of the interviews. These research group members took turns to interview every volunteer and take notes during interviews, following common interview and note-taking protocols (see below).

        \subsubsection{Sample size and composition} The number and characteristics of subjects to be recruited depends on the research goals and the target population. For research on misconceptions, the data analysis plan may involve coding the interview responses and carrying out statistical inference about how often a given misconception occurs, in which case power analysis may be used to choose the sample size.
        
        On the other hand, for validating a concept inventory, a survey questionnaire, or a software product, the purpose of think-alouds is not to estimate a proportion but to find as many as possible of the potential problems in the question wording or the software's usability. For such problem identification studies, especially if budget or time constraints require a small sample size, purposive sampling or targeted recruitment is often seen as appropriate in lieu of random sampling. Researchers may wish to ensure that the interviewees are representative of the target population by including both more-prepared and less-prepared students; different demographic or language groups; or different academic majors or programs.
        \cite{jiyoon2012} administered an early pilot of their concept inventory to a class and used the results to recruit students with diverse scores for cognitive interviews.
        In some cases researchers may also want to compare interviewees from courses with different pedagogical approaches, for instance using traditional vs.\ simulation-based inference, and could keep a record of the textbooks used for each course.
    
        For pretests of survey questionnaires, \citet{Blair_2011} call for larger samples than typical in past practice. In their empirical study on cognitive interviews for a 60-item survey questionnaire, using a total of 5 to 20 interviews would have uncovered only about a quarter to a half of all the wording problems found by using 90 interviews. When improving software usability in an iterative design process, \cite{nielsen1993mathematical} argue for conducting 4--5 think-aloud interviews, using the results to revise the product, and repeating the process many times to identify additional issues. Finally, if the study involves so many questions or tasks that not every interviewee can complete them all, sample sizes should be chosen to ensure adequate coverage per task (see below).
        
        Past statistical concept inventories have reported using small sample sizes and few rounds of question revision: \cite{lane2007development} used two rounds with five and eight students respectively; \cite{jiyoon2012} used two rounds with three and six students respectively; \cite{ziegler2014reconceptualizing} used one round with six students; and \cite{Sabbag:2016} used one round with four students. To ensure confidence that most wording problems can be detected, we encourage future developers of statistics concept inventories to conduct more rounds of wording revisions for a larger total number of interviews. Finally, certain question problems might be more easily detected in some demographic groups than others \citep{Blair_2011}. We encourage inventory developers either to conduct think-alouds with students from a wide range of educational institutions, or to clearly designate a restricted target population for their assessment instrument based on who participated in think-alouds.

        \textbf{In our case}, our goals were exploratory rather than inferential, so we simply interviewed all students who volunteered (22 students in the first semester, three in the next, and six in the last). Because students volunteered to participate, our sample may not be representative of all students who take our introductory courses, though our informal sense was that our interviewees were roughly representative of the demographics of this population.
        We did not record our students' demographics, native language, or major. The introductory statistics course is a requirement for first-year students in the college where our department is located, and students have until the second semester of their sophomore year to declare a major, so many of our students had not yet declared a major. However, this information would be crucial to record and report in studies that wish to generalize beyond the local student population.
        
    \subsection{Conduct interviews} 
    
        \subsubsection{Welcome and introduction} It is important for the subject to feel comfortable during the interview process. As in the recruitment process, power dynamics between the interviewer and interviewee are an important consideration. In ideal circumstances, interviews would be conducted by a non-expert in the subject material to minimize the expert-novice power differential; this is more feasible for think-alouds than for other approaches, such as verbal probing where the interviewer might need domain expertise.
        Regardless, the recruiting script and introductory script should focus on making students as comfortable as possible with the think-aloud process, and interviewers should attempt to present themselves as non-judgmental and supportive throughout the interview process.
        
        The interviewer should begin by welcoming the student; introducing themselves (and the note-taker, if present---see below); and optionally offer the student a bottle of water. At the beginning of the interview, the interviewer explains the interview process and the purpose. As in the recruitment step, it is important to reassure the student that their answers will have no impact on their grade in the course, and that the purpose of the interview is to assess the \textit{course}, \textit{instructor}, and/or \textit{assessment material}, \emph{not} the student.
        
        A sample introduction script can be found in the Supplementary Materials; it is similar to the example language in Table 2.1 of \citet{leighton2017using}. Subjects will also likely need to sign a consent form agreeing to participate in research.
        
        \textbf{In our case}, we did not use non-expert interviewers, as all team members were experts in introductory statistics. Furthermore, as interviews were conducted verbally and in English, non-native speakers may have been less likely to volunteer, or more cautious when voicing their thoughts. Finally, our interviewers were mostly male and/or white, which again could have impacted which students volunteered or how comfortable they felt thinking aloud.
        We attempted to mitigate these concerns through the language in our recruiting and introductory scripts, and through our interviewers' non-judgmental approach to the interview process.
        
        Our introductory script also emphasized that our purpose in investigating student understanding of introductory topics was ultimately to improve our courses, not to evaluate the student.
        
        \subsubsection{Warm-up} Thinking aloud can be challenging, and most subjects don't have experience with this skill. To introduce the idea of thinking aloud, \citet{leighton2017using} and \cite{liu2015overview} recommend a warm-up activity in which the interviewee thinks aloud with a practice problem. Without such practice, students may try to problem-solve first and then justify conclusions out loud afterwards, instead of narrating all along.
        This warm-up should be accessible even without statistical knowledge or, for that matter, cultural knowledge.
        
        \textbf{In our case}, for example, a warm-up used in our interviews was asking the student to describe the steps involved in making their favorite kind of toast. This replaced an initial warm-up activity of discussing a data visualization about an American actor, which turned out to be unnecessarily challenging for novice statistics students as well as for students unfamiliar with US television shows.
        
        \subsubsection{Interview questions} Subjects are given each question in turn, and asked to think aloud while answering. The interviewer does not interrupt, except to remind the interviewee to think aloud if needed. 
        
        The number of interview questions answered by a subject will depend on the length of the questions and the subject's skills. 
        For development of a concept inventory, we recommend varying the question order systematically to ensure equal coverage for all questions. For an exploratory study like ours, question order may be varied to prioritize questions that seem to be provoking rich responses. For a formal study of particular misconceptions, we recommend simply choosing few questions overall, so that every interviewee is likely to complete all tasks.
        
        \textbf{In our case}, our think-aloud interview sessions were structured to include ten minutes for introduction and instructions; about thirty minutes for students to solve questions while thinking aloud uninterrupted; and a twenty-minute period at the end for the interviewer to review the questions with the student, with follow-up discussion to clarify the student’s reasoning as needed, and finally explaining the answers to the student if they should ask.
        
        Our students answered between 6 and 38 questions in the thirty-minute question period, with most students answering about 20. As we drafted more questions than one student could answer in the allotted time, we varied the order in which questions were asked for different students, prioritizing the questions that seemed to be turning up the most interesting responses. As a result, we recorded between 1 and 14 responses for each interview question within each round of interviews, with a mean of 5.4.
        
        \subsubsection{Interview records} While the subject thinks aloud, the interviewer or a second designated note-taker may take notes, including quotes, interesting methods used, and any part of the task the subject found confusing. Alternatively, the interview may be video- or audio-recorded for future analysis. For exploratory think-alouds, note taking may be sufficient to identify broad themes in interviewee responses, and the time cost of transcribing and coding recorded interviews is likely prohibitive. Other research contexts may require careful assessment of each interview (such as detailed coding to count how often particular response strategies were used, or extended quotes to show rich details of interviewee thinking), in which case recording is preferred. If recordings are made, your IRB application will need to explain how you will protect the anonymity and confidentiality of these recordings.
        
        If students use scratch paper while working out their answers, this should also be kept as part of the data for possible analysis.
        
        \textbf{In our case}, our interviews were conducted with one designated interviewer, who sat next to the student and asked questions, and one designated note-taker, who sat at the other end of the room and took notes during the interview process. Both interviewer and note-taker were research group members. Although we did not record interviews, after the first several think-alouds our research team developed a coding structure to help note-takers flag points of interest in real time. For instance, our coding noted when students misunderstood the question or used non-statistical reasoning (question wording or subject matter knowledge) to reach an answer, which helped us flag items that needed to be revised before they could be useful for studying statistical knowledge. Our coding scheme is summarized in the supplementary materials.
        
        \subsubsection{Debrief (student)} To allow the interviewer to ask clarifying questions, time should be allotted for a twenty minute debrief at the end of each interview. Importantly, this also provides an opportunity for the student to ask any questions, and for the interviewer to help the student understand the material better. \cite{leighton2017using} terms this a ``retrospective'' portion of the interview, in contrast to the ``concurrent'' think-aloud portion above. If a note-taker is used, they should clearly delineate which notes come from the concurrent vs.\ retrospective portions.
        
        \textbf{In our case}, we allowed twenty minutes for the debrief.
        
        \subsubsection{Compensation} If possible, interviewees should be compensated for participation in the research process.
        
        \textbf{In our case}, students were given a \$20 Amazon gift card at the end of the interview.
        
        \subsubsection{Debrief (interviewer and note-taker)}  After the interviewee leaves, the interviewer (and note-taker, if present) should take a moment to note any important observations that they did not manage to record during the interview itself.
        
        \textbf{In our case}, the interviewer and note-taker debriefed together.
        This step typically took around five to ten minutes.

    \subsection{Analyze results} If recordings were made, it is generally useful to transcribe the interviews, then code them to show where and how often certain responses occurred. For instance, in an exploratory study on misconceptions or data-analysis practices, initial review of the think-alouds might lead to tabulation of all the strategies that different interviewees used for a task. Each of these strategies might then become a code, and the analysis might involve reflecting on the frequency of each code by task or by sub-groups of interviewees. Meanwhile, for a confirmatory study, the codes should be determined in advance, such as by experts determining a cognitive model of the response processes they expect students to use, along with a rubric for deciding which utterances could count as evidence for or against the model; responses coded by this rubric can be analyzed to determine how well actual student behavior matched the experts' model \citep[Chapter 4]{leighton2017using}. In both cases, most codes will probably need to be task-specific.
    
    However, for developing a concept inventory or a survey, some codes might be reused across tasks, relating to how the items themselves could be improved (e.g., confusing wording; too long; can be answered without statistical knowledge; etc.)\ as well as whether the interviewee's response showed signs of specific expected misconceptions. For instance, \cite{jiyoon2012} coded each response by whether students got the right or wrong answer and also whether they used correct or incorrect reasoning, then reported how often each question had ``matching'' answers (either right and with correct reasoning or wrong and with incorrect reasoning, but not vice versa).
    Extended quotes from the coded transcription can provide detail on exactly what stumbling blocks arose, and may help suggest how to revise the item. In concept inventory writeups, the developers often report each item's original wording, relevant quotes from each interviewee, and consequent changes to the item. If the original item was presented as open-ended, any incorrect responses may be used to develop multiple-choice distractor answers.
    
    To guard against idiosyncratic coding, at least two raters should code several reports using the same coding scheme so that inter-rater reliability can be evaluated. If necessary, rating discrepancies can be reconciled through discussion and the coding scheme can be improved.
    
    As discussed above, \cite{nielsen1993mathematical} recommend frequent iteration cycles of 4--5 interviews followed by revisions. Unless the interview tasks have been extensively pretested already, we suggest planning from the start for at least two cycles of think-alouds---and possibly many more, if the goal is to detect and fix problems with an instrument. The first cycle is likely to find at least some of the most severe problems with the initial tasks or the interview protocol itself; a second cycle at minimum allows researchers to check whether changes to question wording or protocol introduced any new issues. If multiple revision cycles were used, researchers ought to report how they decided when to stop revising.
    
    \textbf{In our case}, our research team met weekly during the first semester of interviews (Spring 2018) and once each during the next two semesters of interviews (Summers 2018 and 2019) to discuss interim results and to propose question revisions or new items.
    We planned to iterate over one or two cycles of small revisions per semester, although for ease of exposition our case studies in Section~\ref{sec:case-studies} focus on scenarios with one major revision each. Furthermore, as we did not anticipate generalizing our results beyond our local student body, we did not plan for recording, transcription, and detailed coding. We found that our note-takers could record the most interesting qualitative takeaways from each interview in adequate detail for our purposes, though such notes may not have been sufficiently detailed or reliable for other research goals. We provide several examples in Section~\ref{sec:case-studies}.

\section{Case studies}
\label{sec:case-studies}

In Section \ref{sec:think-aloud-process}, we described the general think-aloud process, and specific details for our think-aloud interviews with students in introductory statistics courses. Our goal in conducting these interviews was to explore misconceptions in introductory students at our university, and we compiled questions to target misconceptions we had encountered through interactions with students in class, office hours, and assignments. In this section, we describe our experiences with think-aloud interviews for several questions. We focus on five questions in which students produced unexpected answers which revealed misconceptions of which we were previously unaware, and which motivated us to reconsider how we taught these topics. We also take the opportunity to show how an early round of think-alouds can lead to revisions that make the tasks more effective in later think-alouds.

These five questions were tested in think-aloud interviews across 24 different students. We will use numbers 1--17 to denote the students directly quoted or paraphrased in this paper.

\subsection{Sampling distributions and histograms}\label{sec:sampling}

Understanding variability and sampling distributions is an important part of the
GAISE College Report guidelines \citep{gaise2016}, but we have noticed that
students often struggle with these concepts. The introductory statistics course at \blinded{Carnegie Mellon}{our institution} devotes substantial time to sampling distributions, showing students through simulation-based activities how the shape and variance of the sampling distribution of the mean changes as we change the sample size. These activities include sampling from different population distributions, to demonstrate how the central limit theorem applies even when the original distribution is decidedly non-normal.

However, in our experience students often struggle to understand the idea that the variance of the sample mean decreases as sample size increases. To explore student reasoning about variability within sampling distributions and sample size, we drafted a question in which students had to visually identify a decrease in variance, and then connect this with an increase in sample size. However, think-aloud interviews showed that students misinterpreted the histograms we used to display the sampling distributions, and also revealed potential misconceptions about normality of sampling distributions vs.\ normality of the population. This inspired us to revise the original question, draft a new question, and conduct further think-alouds to explore misconceptions.

\subsubsection{Original question} 
\begin{figure}
	\centering
    \includegraphics[width=\textwidth]{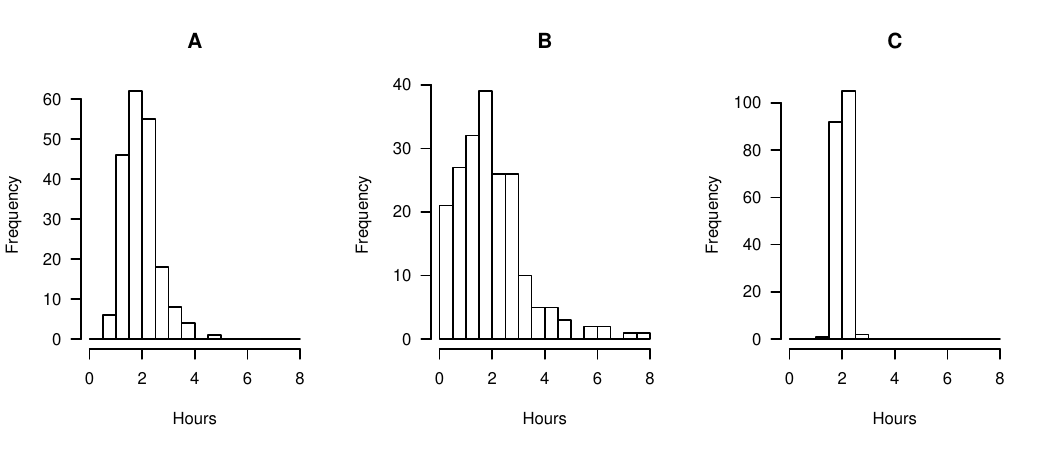}
    \captionsetup{singlelinecheck=off}
    \caption[]{(\qname{study-time}, original version) To estimate the average number of daily hours that students study at a large public college, a researcher randomly samples some students, then calculates the average number of daily study hours for the sample. Pictured (in scrambled order) are three histograms:
One of them represents the population distribution of number of hours studied; the other two are sampling distributions of the average number of hours studied \(\overline{X}\), one
for sample size \(n=5\), and one for sample size \(n=50\). Circle the most likely distribution for each description.

\begin{itemize}
    \item Population distribution: \hspace{14mm} A \hspace{2mm} B \hspace{2mm}  C 
    \item Sampling distribution for $n=5$: \hspace{2mm} A \hspace{2mm} B \hspace{2mm}  C 
    \item Sampling distribution for $n=50$: \hspace{1mm} A \hspace{2mm} B \hspace{2mm}  C 
\end{itemize}}
    \label{fig:study-time}
\end{figure}

Figure \ref{fig:study-time} shows the \qname{study-time} question, intended to test understanding of sampling distributions and sample size. We expected that students who did not understand the relationship between sample size and variance of the sample mean would not know how to choose the correct answer; but they might still get it partially correct if they remembered the approximate normality of the mean's sampling distribution. We were curious to see what other strategies students might use for this problem if they did not recall either of these two concepts. The intended answer was that histogram B is the population distribution, histogram A the sampling distribution of $\overline{X}$ when $n = 5$, and histogram C the sampling distribution with $n = 50$.

\subsubsection{Student responses} To our surprise, all nine students who answered this question during think-aloud interviews got it wrong, claiming that the sampling distribution of the mean with \(n=5\) should be graph C in Figure~\ref{fig:study-time}. No students appeared to use the idea of normality of sampling distributions in their reasoning for this question, and only one student noted that variance should decrease with increasing sample size in a sampling distribution (Student 1). 
No others indicated paying attention to variability. Three students confused the sample size with the number of bars in the histogram, with one student commenting that ``small \(n\) means few bars'' (Student 2)
and then concluding that a sampling distribution with \(n=5\) should have the fewest bars (graph C). Another student admitted, in the retrospective portion of the interview, to not having thought about the sample average at all, just the distribution of the sample (Student 3).
This suggested the question was not capturing the reasoning it was intended to capture: students were selecting histograms by matching \(n\) to the number of bars, not necessarily by reasoning about the variance of the \emph{mean} of samples of varying sizes. This is related to a previously-studied misconception, of which we were unaware, that students mistake the bar heights in a histogram as the observed values in a dataset, and the number of bars as the number of observations \citep{Kaplan:2014vf, boels2019histogramreview}. Additionally, two of the nine students commented that the population should be normally distributed and hence selected graph A as the population distribution, arguing that it was the most symmetric (Students 2 and 4).
Previous research has also identified students incorrectly thinking that distributions besides sampling distributions should have characteristics of the normal distribution \citep{noll2015proper}.

\subsubsection{Revision} Based on these think-aloud results, we took two steps to follow up on the misconceptions that were uncovered. First, would students still fail to relate the spread of the sampling distribution to the sample size if they were not misreading the histograms and statistical jargon? We revised the original \qname{study-time} question to use mechanistic language without mathematical notation, by replacing the initial question text with the following description and asking students to match A, B, and C to Jeri, Steve, and Cosma, keeping the figure the same:
\begin{quote}
    Jeri, Steve, and Cosma are conducting surveys of how many hours students study per day at a
large public university.

    Jeri talks to two hundred students, \textbf{one at a time}, and adds each student’s answer to her
histogram.

    Steve talks to two hundred~\textbf{groups of 5 students}. After asking each group of 5 students how
much they study, Steve takes the~\textbf{group's average} and adds it to his histogram.

  Cosma talks to two hundred~\textbf{groups of 50 students}. After asking
  each group of 50 students how much they study, Cosma takes
  the \textbf{group's average} and adds it to his histogram.
  
  The three final histograms are shown below, in scrambled order.
\end{quote}

Because the number of points in each histogram---two hundred---was explicitly
stated in each case, we hoped that students would no longer answer incorrectly due to misreading the histograms. This version
also does not use the term ``sampling distribution'', so it tests whether students recognize the concept without seeing the term.

\begin{figure}
    \centering
    A: \raisebox{-0.5\height}{\includegraphics[width=0.8\textwidth]{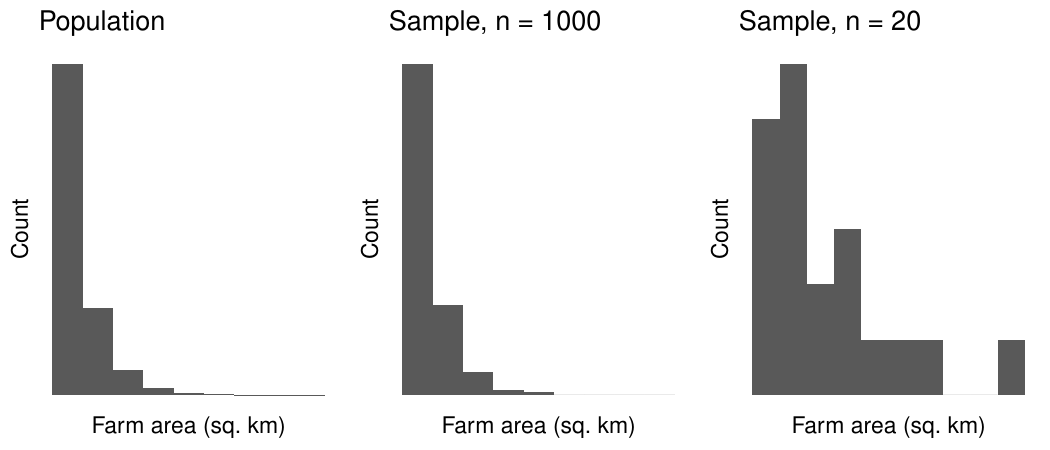}}

    \hrulefill
    
    B: \raisebox{-0.5\height}{\includegraphics[width=0.8\textwidth]{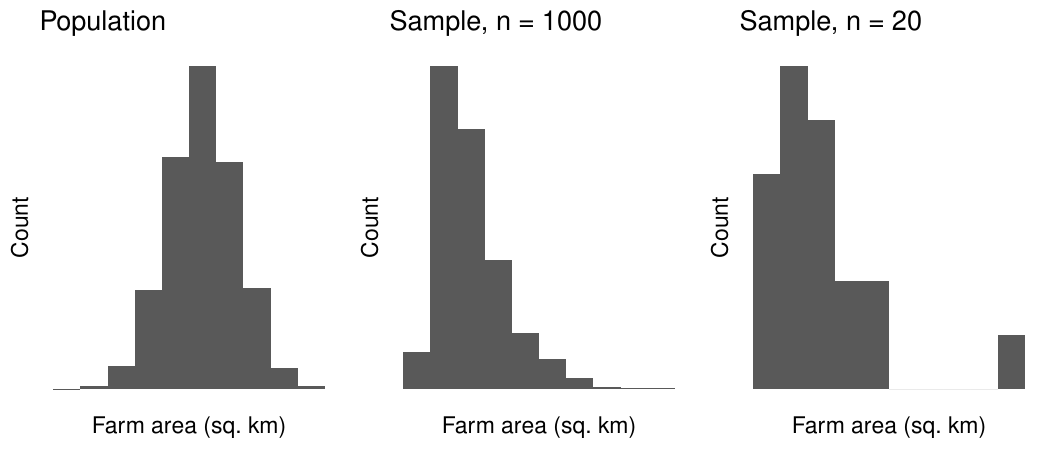}}

    \hrulefill

    C: \raisebox{-0.5\height}{\includegraphics[width=0.8\textwidth]{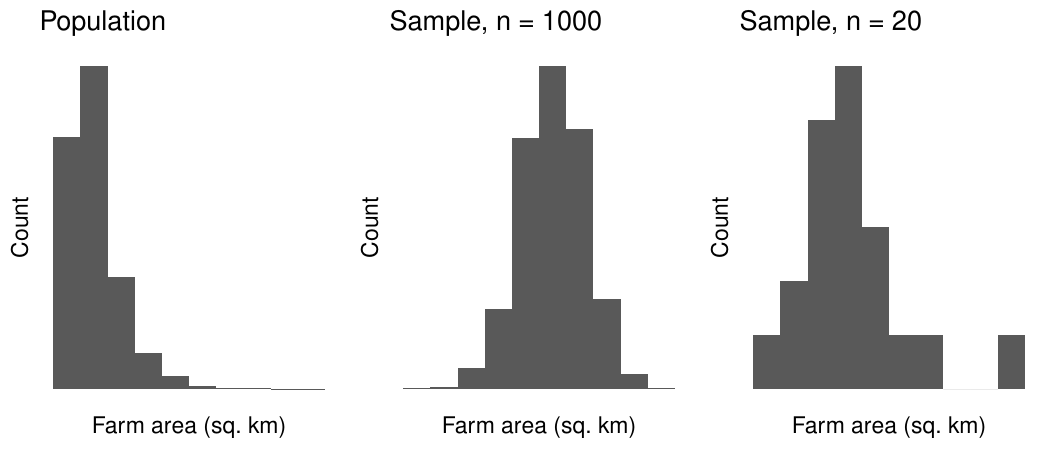}}
\caption{ (\qname{farm-areas}) Farmer Brown collects data on the land area of farms in the US (in square
kilometers). By surveying her farming friends, she collects the area of every
farm in the US, and she makes a histogram of the population distribution of US
farm areas. She then takes two random samples from the population, of sizes $n =
1000$ and $n = 20$, and plots histograms of the values in each sample. One of the rows below shows her three histograms. Using the \textbf{shape} of
the histograms, choose the correct row.}
    \label{fig:farm-areas}
\end{figure}

Second, we also drafted a new question to further explore the potential misconception that populations are always normally distributed. Would students still have this misconception when we are not directly asking about the tricky topic of sampling distributions? The \qname{farm-areas} question, shown in Figure~\ref{fig:farm-areas}, describes a situation in which the entire population is surveyed, and a histogram of the results prepared, along with histograms of samples---\emph{not} sampling distributions---of sizes \(n = 20\) and \(n = 1000\). Three possible sets of histograms are provided, and students are asked to select the most plausible set based on their shapes. The intended answer, (A), shows a skewed population distribution and two skewed samples. The first distractor, (B), is meant to test whether students are willing to believe that a population could be normally distributed even if a large sample has a skewed distribution. The second distractor, (C), was included to test the opposite misconception: that the distribution of a sample would appear normal, even if the population does not. We expected students to choose answer (C) if they confused the distribution of a sample with the sampling distribution \citep{lipson2002role, Chance:2004tt, Castro_Sotos_2007, Kaplan:2014vf}.

\subsubsection{More student responses} In twelve new think-aloud interviews on the revised \qname{study-time} question, nine students answered correctly. However, three of those nine still confused the number of bars with the sample size, as did one student who answered incorrectly. These four students misread the text and thought that there were 200 students total, so that Cosma had four groups of 50. When combined with the histogram-bars-as-data-points misconception, they correctly matched Cosma to graph C despite making two major mistakes in reasoning.
Another student who answered incorrectly did use correct reasoning about the normality but not the spread of sampling distributions; they wrongly matched Cosma's larger groups of students with graph A because it looked more normal (Student 5).
Of the remaining correct answers, five students referenced the
normality or spread of the distribution of means, saying things like ``taking
the average of a larger group should lead to the means being all bunched up in
one place'' (Student 6).
In short, more students did appear to use some of the intended reasoning in answering this question than in its original version, although this question would benefit from further rounds of revision.
As with other misinterpretations of histograms that have been previously reported in the literature
\citep{Kaplan:2014vf,Cooper_2018,Cooper:2008up}, students continued to misinterpret the meaning of histogram bars.

Ten students answered the \qname{farm-areas} question during think-aloud interviews, of whom only four selected the intended answer.
The remaining six split evenly between the two distractor answers, reinforcing the notion that some of our students do hold misconceptions about normality of populations and about samples vs.\ sampling distributions. Among those selecting the first distractor (row B in Figure \ref{fig:farm-areas}), one explained that with a larger sample size, ``there is less of a chance for data to vary'' (Student 7),
and the distractor had the most ``centralized'' population distribution.
In the retrospective portion of the interview, the student confirmed that this meant they had been expecting to see a symmetric population distribution.
Among students selecting the second distractor (row C in Figure \ref{fig:farm-areas}), one noted ``I'm assuming it's looking for a normal distribution, the greater the sample size'' (Student 8)
and indicated that the choice had a more normal histogram for $n = 1000$, suggesting that they were indeed looking for the normality that would be expected if these were sampling distributions rather than samples.

\subsubsection{Discussion} In this case, think-aloud interviews allowed us to identify misconceptions we were unaware of, and draft some new materials to further explore these misconceptions. These exploratory results, however, do not by themselves explain why students hold these misconceptions, and it is unclear whether misunderstandings arise due to the way histograms and sampling distributions are presented in our statistics courses. Further research could use think-alouds as one tool to explore how students think about sampling, perhaps in conjunction with specific teaching interventions. In the short term, we have begun to directly address these misconceptions when teaching students about histograms and about the distinctions between populations, samples, and sampling distributions.

The questions and graphs presented here are by no means fully polished, and additional think-alouds could be used to further improve and refine them. For instance, the lack of marked x- and y-axis scales in \qname{farm-areas} may have introduced new confusion---distinct from the histogram-reading difficulties we already uncovered---that should be addressed in future rounds of revisions and new think-alouds. However, even in unpolished form, these questions have proved useful for our purposes of investigating our students' understanding.

\subsection{Correlation and causation}

The role of random assignment in drawing causal conclusions is emphasized by the GAISE guidelines, under the goal that students should be able to ``explain the central role of randomness in designing studies and drawing conclusions'' \citep{gaise2016}. Our introductory courses have therefore emphasized the difference between randomized experiments and observational studies, and that correlation does not necessarily imply causation. Activities include examples of data analyses in which students critique the language used to discuss causation vs.\ observation, and identify instances in which causal conclusions have been incorrectly drawn.

For think-aloud interviews, we drafted two questions on correlation and causation, based on our own class materials. However, think-aloud interviews suggested that some students were unwilling to \textit{ever} draw causal conclusions, a misconception we targeted with a new question in a second round of interviews.

\subsubsection{Initial questions}  In \qname{clinical-trial}, a randomized experiment supports a causal conclusion, while in \qname{books}, an observational study does not support a causal claim. Table \ref{tab:correlation-causation-qs} shows the initial questions and answer choices. We expected that among our students who had just begun learning about these topics, the most common mistake would be the one that our courses usually try to prevent: making causal claims where they are not warranted (in the \qname{books} question).

\begin{table}
\begin{center}
\footnotesize
\begin{tabular}{|p{6in}|} \hline 
\qname{clinical-trial} (original) \vskip 2mm
A clinical trial randomly assigned subjects to receive either vitamin C or a placebo as a treatment for a cold. The trial found a statistically significant negative correlation between vitamin C dose and the duration of cold symptoms.
\vskip 1mm
Which of the following can we conclude?
\vskip 1mm
A. Recovering faster from a cold causes subjects to take more vitamin C.\\
B. Taking more vitamin C causes subjects to recover faster from a cold.\\
C. We cannot draw any conclusions because correlation does not imply causation.\\
D. We cannot draw any conclusions because assignment was random instead of systematic.

\\ \hline 
\qname{books} \vskip 2mm
A survey of Californians found a statistically significant positive correlation
between number of books read and nearsightedness.
\vskip 1mm
Which of the following can we conclude about Californians?
\vskip 1mm   
A. Reading books causes an increased risk of being nearsighted.\\
B. Being nearsighted causes people to read more books.\\
C. We cannot determine which factor causes the other, because correlation does not imply causation.\\
D. We cannot draw any conclusions because Californians aren't a random sample of people.
\\ \hline 
\end{tabular} 
\end{center} 
\caption{Initial questions, with answer choices, on correlation and causation.}
\label{tab:correlation-causation-qs}
\end{table}

\subsubsection{Student responses}  For \qname{clinical-trial}, the intended answer choice (B) is that vitamin C causes faster recovery from colds,  because the study described is a randomized experiment. In think-aloud interviews, four of six students answered correctly; however, none of these four students referred to random assignment as they thought aloud. Two students who answered correctly told us they strongly believed that ``correlation does not equal causation,'' but still picked the intended answer because it made sense to them that vitamin C actually would cause subjects to recover faster from a cold (Students 9 and 10).
One said you ``usually can't assume causation'' (Student 9),
then picked the causal answer despite hesitating and stating that it is just correlation. While students may get questions (particularly multiple choice questions) correct for the wrong reason, or just by guessing, in initial think-alouds with \qname{clinical-trial} we saw students answering the question correctly despite truly believing the opposite conclusion (that ``correlation does not equal causation''). Furthermore, of the two who answered incorrectly, both chose answer C, refusing to make causal claims despite the random assignment. One believed that you can only ever talk about significance, not causation (Student 11),
while the other stated they did not see any difference between this question and \qname{books} (Student 12).

On the other hand, in think-alouds for \qname{books}, four of five students chose the intended answer, (C): ``We cannot determine which factor causes the other, because correlation does not imply causation.'' In their responses, students said nothing to indicate they understood when causal conclusions could actually be drawn; one student explicitly stated a misconception that ``correlation does not imply causation is a universal rule'' (Student 11).
The fifth student, who answered incorrectly, tried to use elimination instead of statistical reasoning (Student 13).

\subsubsection{Revision} Student responses to \qname{books} and \qname{clinical-trial} suggest that our students were generally overcautious about drawing causal conclusions. They clung to the mantra ``correlation is not causation'' and based their causal claims not on statistical grounds of study design but on subject-matter plausibility. From this original pair of questions, we could not tell whether confusion about causation arose mostly because they were primed by this mantra, or whether students truly misunderstood the roles of random assignment and confounding variables in making causal claims.

We therefore drafted an additional question, \qname{font-test} (Table \ref{tab:correlation-causation-qs-new}),  which explicitly described a randomized experiment and included distractor answers focusing on confounding variables and random assignment---but using mechanistic language that was intended to avoid the technical terms ``correlation'' and ``causation'' (as well as ``statistically significant''). The intended answer was (A), while we expected students who misunderstood confounding or study design to select answers (C) or (D). Additionally, we changed the treatment in \qname{clinical-trial} from vitamin C to mindfulness meditation, so that the treatment's efficacy would seem like more of an open question. We also softened the wording of the question from ``Which of the following can we conclude?'' to ``Which conclusion does this support?''

\begin{table}
\begin{center} 
\footnotesize
\begin{tabular}{|p{6in}|} \hline 
\qname{clinical-trial} (revised) \vskip 2mm
A clinical trial randomly assigned subjects to either practice mindfulness meditation or a
placebo relaxation exercise as a treatment for a cold. The trial found that subjects who
practiced mindfulness meditation had a shorter time to recovery than students assigned to the
relaxation exercise, and the result was statistically significant.
\vskip 1mm
Which conclusion does this support?
\vskip 1mm
A. Recovering faster from a cold causes subjects to meditate.\\
B. Mindfulness meditation causes subjects to recover faster from a cold.\\
C. We cannot draw any conclusions because correlation does not imply causation.\\
D. We cannot draw any conclusions because assignment was random instead of systematic.

\\ \hline 
\qname{font-test} \vskip 2mm
Professor Smith wants to know if typing her introductory statistics exams in
Comic Sans will improve their exam performance. To answer this question, she
randomly gives half of the 200 students in her class an exam with all of the
questions typed in Comic Sans, while the other students get the same exam with
questions typed in Times New Roman.

After comparing the exam scores across both groups of students, Professor Smith
finds that the students who were given the exam typed in Comic Sans had a higher
average grade on the exam, compared to the average grade for students who did
not receive the exam typed in Comic Sans. Professor Smith repeats this
experiment across multiple semesters of her course and always sees the same
result.

\vskip 1mm  
Which of the following is true?
\vskip 1mm     
A. The result is statistical evidence that giving students exams typed in Comic Sans will lead to higher exam scores across the class.\\
B. All teachers in every subject should print their exams in Comic Sans to improve their students' performance.\\
C. Professor Smith can't draw any conclusions from these tests because other factors, such as the amount of hours students spent studying, might also affect their exam results.\\
D. Professor Smith can't draw any conclusions from these tests because she randomly decided which students would receive the exam typed in Comic Sans instead of choosing students systematically, such as giving only the female students the exam typed in Comic Sans.
\\ \hline \end{tabular} \end{center} \caption{Revised and new questions, with answer choices, about correlation and causation.}
\label{tab:correlation-causation-qs-new}
\end{table}

\subsubsection{More student responses} The incorrect responses to \qname{font-test} (three of six correct) indicated misunderstandings beyond mere reliance on the mantra ``correlation is not causation'': one student conflated random sampling with random assignment (Student 14),
and another thought of possible confounding factors and did not notice the random assignment (Student 15),
while the third noticed random assignment but still thought there were confounding variables (Student 16).
Of the three correct responses, two showed an understanding that random assignment allows for causal claims, but the third wrongly justified the causal claim with random sampling (Student 17). 
For the revised \qname{clinical-trial}, all five students incorrectly answered that correlation does not imply causation, which reinforced our suspicion that correct answers to the original version stemmed from the plausibility of vitamin C treatment rather than from understanding statistical concepts. One student thought there could still be confounding variables (Student 16),
while another seemed to believe that correlation could never imply causation (Student 17), 
a belief also expressed in response to \qname{font-test}:
\begin{itemize}
    \item ``When can we ever say something causes something else?'' (\qname{font-test}, Student 14)
    \item ``I think the word `causes' is too strong... my friend who's a stats major always tells me you can't say this causes that---there's always other factors'' (\qname{clinical-trial}, Student 14)
\end{itemize}

\subsubsection{Discussion} It seems that while our students had learned that correlation does not imply causation, they struggled more with understanding how randomized experiments could provide evidence for causal conclusions. This matches with several studies in the statistics education literature: \cite{pfannkuch2015experiment} and \cite{sawilowsky2004teaching} observed that students believed confounding was possible even in randomized experiments, while \cite{fry2017introductory} discusses the misconception of ``not believing causal claims can be made even though random assignment was used.''

We hypothesize that an incomplete understanding of confounding variables, and why randomized experiments prevent confounding, may be part of the confusion with drawing causal conclusions. If this is the case, it may be helpful to include additional causal inference material in the introductory curriculum; inspired by these think-alouds, and the work of \cite{Cummiskey:2020} and \cite{lubke2020causal}, we have introduced simple causal diagrams into an introductory statistics course. Some preliminary discussion on this new material can be found in \blinded{\cite{evans2020ecots}}{[citation blinded]}. It is also possible that students were confused because our courses have overemphasized observational studies and under-emphasized randomized experiments, or simply because of question wording---as in Section~\ref{sec:sampling}, the questions presented here could be improved by further think-alouds. In future work, we hope to further explore why our students hesitate with causation.

\section{Conclusion}

In our experiences with think-aloud interviews, we have seen that think-alouds provided a valuable tool for investigating student understanding of introductory statistics concepts. By conducting interviews with students in our own courses, we learned that we had not adequately anticipated certain misconceptions about histograms, sampling distributions, and correlation and causation. Our findings so far have inspired us to plan for future think-aloud interviews where we will further explore our students' reasoning about study design, data analysis, and statistical inference. For example, we hope to conduct future interviews in which students conduct or assess data analysis tasks, to see which choices students make (and in what order) when working with data. Many of the steps from Section \ref{sec:think-aloud-process} would be similar, but we would need to carefully choose questions that provide enough structure that they can be completed during an interview, while still allowing students to make different choices.

The way we designed our existing think-aloud study was suited to our particular needs.
The real-time nature of think-alouds allowed us to gauge how well students' statistical thinking had become internalized, rather than limited to the more deliberate, self-conscious reflection we would have seen with verbal probing or during office hours.
By using a process with more than one iteration---conduct several think-alouds, reflect on student responses, revise questions or draft new ones, and repeat---we were able to adapt quickly and follow up on surprising findings, unlike with a static concept inventory.
Finally, as a research group composed of instructors with a common student population, our shared discussions of student responses prompted buy-in to making changes to our own courses, including new material designed to address the misconceptions we were seeing.
Of course, we present this as just one example of implementing think-alouds, and other situations will call for a different approach.

We hope that our experiences encourage other statistics education researchers to use think-aloud interviews, whether they are investigating misconceptions, writing questions to assess a single concept, or revising a full concept inventory. Likewise, we hope that our summary of best practices will help others tailor their own think-aloud study designs to their institutional contexts and research problems.

\section*{Acknowledgments}

\if@anonymize
\else
We are grateful to the editor, associate editors, and reviewers for their many helpful comments. Thanks to the Carnegie Mellon's Eberly Center for Teaching Excellence and Educational Innovation for initial support developing this study; to David Gerritsen for initial advice on conducting think-aloud interviews;
and to Gordon Weinberg for feedback and suggestions for questions, and for facilitating administration of the assessment to his courses. We also thank Sangwon Hyun, Ron Yurko, and Kevin Lin for contributing questions and assisting with think-aloud interviews. We are grateful to Christopher Peter Makris for extensive logistical support. Many thanks to our student participants, without whom this research would not have been possible.
\fi

\appendix

\section*{Supplemental Materials}

The recruiting script, interview protocol, and coding scheme are included as supplemental material.


\bibliography{think-aloud-refs}

\begin{thebibliography}{}

\bibitem[Adams and Wieman, 2011]{Adams2011}
Adams, W.~K. and Wieman, C.~E. (2011).
\newblock {Development and Validation of Instruments to Measure Learning of
  Expert-Like Thinking}.
\newblock {\em International Journal of Science Education}, 33(9):1289--1312.

\bibitem[Bandalos, 2018]{Bandalos:2018}
Bandalos, D.~L. (2018).
\newblock {\em Measurement Theory and Applications for the Social Sciences}.
\newblock Guildford Press, New York, NY.

\bibitem[Blair and Conrad, 2011]{Blair_2011}
Blair, J. and Conrad, F.~G. (2011).
\newblock Sample size for cognitive interview pretesting.
\newblock {\em Public Opinion Quarterly}, 75(4):636--658.

\bibitem[Boels et~al., 2019]{boels2019histogramreview}
Boels, L., Bakker, A., {Van Dooren}, W., and Drijvers, P. (2019).
\newblock Conceptual difficulties when interpreting histograms: A review.
\newblock {\em Educational Research Review}, 28:100291.

\bibitem[Bowen, 1994]{Bowen_1994}
Bowen, C.~W. (1994).
\newblock {Think-Aloud Methods in Chemistry Education: Understanding Student
  Thinking}.
\newblock {\em Journal of Chemical Education}, 71(3):184.

\bibitem[Branch, 2000]{branch2000investigating}
Branch, J.~L. (2000).
\newblock Investigating the information-seeking processes of adolescents: The
  value of using think alouds and think afters.
\newblock {\em Library \& Information Science Research}, 22(4):371--392.

\bibitem[Castro~Sotos et~al., 2007]{Castro_Sotos_2007}
Castro~Sotos, A.~E., Vanhoof, S., Van~den Noortgate, W., and Onghena, P.
  (2007).
\newblock Students' misconceptions of statistical inference: A review of the
  empirical evidence from research on statistics education.
\newblock {\em Educational Research Review}, 2(2):98--113.

\bibitem[Chance et~al., 2004]{Chance:2004tt}
Chance, B., delMas, R., and Garfield, J. (2004).
\newblock Reasoning about sampling distributions.
\newblock In Ben-Zvi, D. and Garfield, J., editors, {\em The Challenge of
  Developing Statistical Literacy, Reasoning and Thinking}, chapter~13, pages
  295--323. Kluwer Academic Publishers.

\bibitem[Cooper, 2018]{Cooper_2018}
Cooper, L.~L. (2018).
\newblock Assessing students' understanding of variability in graphical
  representations that share the common attribute of bars.
\newblock {\em Journal of Statistics Education}, 26(2):110--124.

\bibitem[Cooper and Shore, 2008]{Cooper:2008up}
Cooper, L.~L. and Shore, F.~S. (2008).
\newblock Students' misconceptions in interpreting center and variability of
  data represented via histograms and stem-and-leaf plots.
\newblock {\em Journal of Statistics Education}, 16(2).

\bibitem[Cummiskey et~al., 2020]{Cummiskey:2020}
Cummiskey, K., Adams, B., Pleuss, J., Turner, D., Clark, N., and Watts, K.
  (2020).
\newblock Causal inference in introductory statistics courses.
\newblock {\em Journal of Statistics Education}, 28(1):2--8.

\bibitem[Deane et~al., 2014]{Deane:2014}
Deane, T., Nomme, K., Jeffery, E., Pollock, C., and Birol, G. (2014).
\newblock Development of the {B}iological {E}xperimental {D}esign {C}oncept
  {I}nventory ({BEDCI}).
\newblock {\em CBE—Life Sciences Education}, 13(3):540--551.

\bibitem[Ericsson and Simon, 1998]{Ericsson_1998}
Ericsson, K.~A. and Simon, H.~A. (1998).
\newblock {How to Study Thinking in Everyday Life: Contrasting Think-Aloud
  Protocols With Descriptions and Explanations of Thinking}.
\newblock {\em Mind Culture, and Activity}, 5(3):178--186.

\bibitem[Evans et~al., 2020]{evans2020ecots}
Evans, C., Reinhart, A., Burckhardt, P., Nugent, R., and Weinberg, G. (2020).
\newblock Exploring how students reason about correlation and causation.
\newblock
  \textsc{url:}~\url{https://www.causeweb.org/cause/ecots/ecots20/posters/2-03}.
\newblock Poster presented at: Electronic Conference On Teaching Statistics
  (eCOTS).

\bibitem[Feldon, 2007]{Feldon_2006}
Feldon, D.~F. (2007).
\newblock The implications of research on expertise for curriculum and
  pedagogy.
\newblock {\em Educational Psychology Review}, 19(2):91--110.

\bibitem[Fry, 2017]{fry2017introductory}
Fry, E. (2017).
\newblock {\em Introductory statistics students’ conceptual understanding of
  study design and conclusions}.
\newblock PhD thesis, University of Minnesota.

\bibitem[{GAISE College Report ASA Revision Committee}, 2016]{gaise2016}
{GAISE College Report ASA Revision Committee} (2016).
\newblock {Guidelines for Assessment and Instruction in Statistics Education
  College Report}.
\newblock
  https://www.amstat.org/education/guidelines-for-assessment-and-instruction-in-statistics-education-(gaise)-reports.

\bibitem[Garvin-Doxas and Klymkowsky, 2008]{GarvinDoxas_2008}
Garvin-Doxas, K. and Klymkowsky, M.~W. (2008).
\newblock Understanding randomness and its impact on student learning: Lessons
  learned from building the {B}iology {C}oncept {I}nventory ({BCI}).
\newblock {\em CBE—Life Sciences Education}, 7(2):227--233.

\bibitem[Jorion et~al., 2015]{Jorion_2015}
Jorion, N., Gane, B.~D., James, K., Schroeder, L., DiBello, L.~V., and
  Pellegrino, J.~W. (2015).
\newblock An analytic framework for evaluating the validity of concept
  inventory claims.
\newblock {\em Journal of Engineering Education}, 104(4):454--496.

\bibitem[Kaczmarczyk et~al., 2010]{Kaczmarczyk:2010}
Kaczmarczyk, L.~C., Petrick, E.~R., East, J.~P., and Herman, G.~L. (2010).
\newblock Identifying student misconceptions of programming.
\newblock In {\em Proceedings of the 41st ACM Technical Symposium on Computer
  Science Education}, SIGCSE '10, page 107–111, New York, NY, USA.
  Association for Computing Machinery.

\bibitem[Kaplan et~al., 2014]{Kaplan:2014vf}
Kaplan, J.~J., Gabrosek, J.~G., Curtiss, P., and Malone, C. (2014).
\newblock Investigating student understanding of histograms.
\newblock {\em Journal of Statistics Education}, 22(2).

\bibitem[Karpierz and Wolfman, 2014]{Kuba:2014}
Karpierz, K. and Wolfman, S.~A. (2014).
\newblock Misconceptions and concept inventory questions for binary search
  trees and hash tables.
\newblock In {\em Proceedings of the 45th ACM Technical Symposium on Computer
  Science Education}, SIGCSE '14, page 109–114, New York, NY, USA.
  Association for Computing Machinery.

\bibitem[Konold, 1989]{konold1989informal}
Konold, C. (1989).
\newblock Informal conceptions of probability.
\newblock {\em Cognition and instruction}, 6(1):59--98.

\bibitem[Lane-Getaz, 2007]{lane2007development}
Lane-Getaz, S.~J. (2007).
\newblock {\em Development and Validation of a Research-based Assessments:
  Reasoning about P-values and Statistical Significance}.
\newblock PhD thesis, University of Minnesota.

\bibitem[Leighton, 2013]{leighton2013item}
Leighton, J.~P. (2013).
\newblock Item difficulty and interviewer knowledge effects on the accuracy and
  consistency of examinee response processes in verbal reports.
\newblock {\em Applied Measurement in Education}, 26(2):136--157.

\bibitem[Leighton, 2017]{leighton2017using}
Leighton, J.~P. (2017).
\newblock {\em Using Think-Aloud Interviews and Cognitive Labs in Educational
  Research}.
\newblock Oxford University Press.

\bibitem[Leighton, 2021]{leighton2021rethinking}
Leighton, J.~P. (2021).
\newblock Rethinking think-alouds: The often-problematic collection of response
  process data.
\newblock {\em Applied Measurement in Education}, 34(1):61--74.

\bibitem[Lipson, 2002]{lipson2002role}
Lipson, K. (2002).
\newblock The role of computer based technology in developing understanding of
  the concept of sampling distribution.
\newblock In {\em Proceedings of the Sixth International Conference on Teaching
  Statistics}.

\bibitem[Liu and Li, 2015]{liu2015overview}
Liu, P. and Li, L. (2015).
\newblock An overview of metacognitive awareness and l2 reading strategies.
\newblock In Wegerif, R., Li, L., and Kaufman, J.~C., editors, {\em The
  Routledge International Handbook of Research on Teaching Thinking},
  chapter~22, pages 290--303. Routledge.

\bibitem[Lovett, 2001]{lovett2001collaborative}
Lovett, M. (2001).
\newblock A collaborative convergence on studying reasoning processes: A case
  study in statistics.
\newblock In Carver, S.~M. and Klahr, D., editors, {\em Cognition and
  Instruction: Twenty-five Years of Progress}, chapter~11, pages 347--384.
  Lawrence Erlbaum Associates Publishers.

\bibitem[L{\"u}bke et~al., 2020]{lubke2020causal}
L{\"u}bke, K., Gehrke, M., Horst, J., and Szepannek, G. (2020).
\newblock Why we should teach causal inference: Examples in linear regression
  with simulated data.
\newblock {\em Journal of Statistics Education}, 28(2):133--139.

\bibitem[McGinness and Savage, 2016]{McGinness_2016}
McGinness, L.~P. and Savage, C.~M. (2016).
\newblock Developing an action concept inventory.
\newblock {\em Physical Review Physics Education Research}, 12:010133.

\bibitem[Meyer et~al., 2020]{meyer2020using}
Meyer, M., Orellana, J., and Reinhart, A. (2020).
\newblock Using cognitive task analysis to uncover misconceptions in
  statistical inference courses.
\newblock
  \textsc{url:}~\url{https://www.causeweb.org/cause/ecots/ecots20/posters/2-02}.
\newblock Poster presented at: Electronic Conference On Teaching Statistics
  (eCOTS).

\bibitem[Newman et~al., 2016]{Newman_2016}
Newman, D.~L., Snyder, C.~W., Fisk, J.~N., and Wright, L.~K. (2016).
\newblock Development of the central dogma concept inventory ({CDCI})
  assessment tool.
\newblock {\em CBE—Life Sciences Education}, 15(2).

\bibitem[Nielsen and Landauer, 1993]{nielsen1993mathematical}
Nielsen, J. and Landauer, T.~K. (1993).
\newblock A mathematical model of the finding of usability problems.
\newblock In {\em Proceedings of the INTERACT '93 and CHI '93 Conference on
  Human Factors in Computing Systems}, CHI '93, page 206–213, New York, NY,
  USA. Association for Computing Machinery.

\bibitem[Noll and Hancock, 2015]{noll2015proper}
Noll, J. and Hancock, S. (2015).
\newblock Proper and paradigmatic metonymy as a lens for characterizing student
  conceptions of distributions and sampling.
\newblock {\em Educational Studies in Mathematics}, 88(3):361--383.

\bibitem[N{\o}rgaard and Hornb{\ae}k, 2006]{Norgaard:2006}
N{\o}rgaard, M. and Hornb{\ae}k, K. (2006).
\newblock What do usability evaluators do in practice?: an explorative study of
  think-aloud testing.
\newblock In {\em {Proceedings of the 6th Conference on Designing Interactive
  Systems}}, pages 209--218.

\bibitem[Park, 2012]{jiyoon2012}
Park, J. (2012).
\newblock {\em {Developing and validating an instrument to measure college
  students' inferential reasoning in statistics: an argument-based approach to
  validation}}.
\newblock PhD thesis, University of Minnesota.

\bibitem[Pfannkuch et~al., 2015]{pfannkuch2015experiment}
Pfannkuch, M., Budgett, S., and Arnold, P. (2015).
\newblock Experiment-to-causation inference: Understanding causality in a
  probabilistic setting.
\newblock In Zieffler, A. and Fry, E., editors, {\em Reasoning about
  Uncertainty: Learning and Teaching Informal Inferential Reasoning},
  chapter~4, pages 95--127. Catalyst Press.

\bibitem[Porter et~al., 2019]{Porter_2019}
Porter, L., Zingaro, D., Liao, S.~N., Taylor, C., Webb, K.~C., Lee, C., and
  Clancy, M. (2019).
\newblock {BDSI}: A validated concept inventory for basic data structures.
\newblock In {\em Proceedings of the 2019 ACM Conference on International
  Computing Education Research}, ICER '19, pages 111--119, New York, NY, USA.
  Association for Computing Machinery.

\bibitem[Pressley and Afflerbach, 1995]{pressley2012verbal}
Pressley, M. and Afflerbach, P. (1995).
\newblock {\em Verbal protocols of reading: The nature of constructively
  responsive reading}.
\newblock Routledge.

\bibitem[Roberts and Fels, 2006]{roberts2006methods}
Roberts, V.~L. and Fels, D.~I. (2006).
\newblock Methods for inclusion: Employing think aloud protocols in software
  usability studies with individuals who are deaf.
\newblock {\em International Journal of Human-Computer Studies},
  64(6):489--501.

\bibitem[Sabbag, 2016]{Sabbag:2016}
Sabbag, A. (2016).
\newblock {\em Examining The Relationship Between Statistical Literacy And
  Statistical Reasoning}.
\newblock PhD thesis, University of Minnesota.

\bibitem[Sawilowsky, 2004]{sawilowsky2004teaching}
Sawilowsky, S.~S. (2004).
\newblock Teaching random assignment: do you believe it works?
\newblock {\em Journal of Modern Applied Statistical Methods}, 3(1):221--226.

\bibitem[Taylor et~al., 2020]{Taylor:2020}
Taylor, C., Clancy, M., Webb, K.~C., Zingaro, D., Lee, C., and Porter, L.
  (2020).
\newblock The practical details of building a cs concept inventory.
\newblock In {\em Proceedings of the 51st ACM Technical Symposium on Computer
  Science Education}, SIGCSE '20, page 372–378, New York, NY, USA.
  Association for Computing Machinery.

\bibitem[Theobold, 2021]{theobold2021oral}
Theobold, A.~S. (2021).
\newblock Oral exams: A more meaningful assessment of students’
  understanding.
\newblock {\em Journal of Statistics and Data Science Education},
  29(2):156--159.

\bibitem[Williams, 1999]{williams1999novice}
Williams, A.~M. (1999).
\newblock Novice students' conceptual knowledge of statistical hypothesis
  testing.
\newblock In Truran, J.~M. and Truran, K.~M., editors, {\em Making the
  difference: Proceedings of the Twenty-second Annual Conference of the
  Mathematics Education Research Group of Australasia}, pages 554--560.
  Adelaide, South Australia: MERGA.

\bibitem[Willis, 2005]{Willis_2005}
Willis, G.~B. (2005).
\newblock {\em Cognitive Interviewing}.
\newblock SAGE Publications.

\bibitem[Woodard and Lee, 2021]{woodard_2021}
Woodard, V. and Lee, H. (2021).
\newblock How students use statistical computing in problem solving.
\newblock {\em Journal of Statistics and Data Science Education},
  29(sup1):S145--S156.

\bibitem[Wren and Barbera, 2013]{Wren_2013}
Wren, D. and Barbera, J. (2013).
\newblock Gathering evidence for validity during the design, development, and
  qualitative evaluation of thermochemistry concept inventory items.
\newblock {\em Journal of Chemical Education}, 90(2):1590--1601.

\bibitem[Ziegler, 2014]{ziegler2014reconceptualizing}
Ziegler, L.~A. (2014).
\newblock {\em Reconceptualizing statistical literacy: Developing an assessment
  for the modern introductory statistics course}.
\newblock PhD thesis, University of Minnesota.

\end{thebibliography}

\end{document}